\documentclass[english,twocolumn]{revtex4}
\usepackage[T1]{fontenc}
\usepackage[latin9]{inputenc}
\usepackage{xcolor}
\usepackage{pdfcolmk}
\usepackage{amsmath}
\usepackage{amssymb}
\usepackage{esint}
\PassOptionsToPackage{normalem}{ulem}
\usepackage{ulem}

\makeatletter

\providecolor{lyxadded}{rgb}{0.99,0.0078,1}
\providecolor{lyxdeleted}{rgb}{1,0,0}

\DeclareRobustCommand{\lyxsout}[1]{\ifx\\#1\else\sout{#1}\fi}

\@ifundefined{textcolor}{}
{%
 \definecolor{BLACK}{gray}{0}
 \definecolor{WHITE}{gray}{1}
 \definecolor{RED}{rgb}{1,0,0}
 \definecolor{GREEN}{rgb}{0,1,0}
 \definecolor{BLUE}{rgb}{0,0,1}
 \definecolor{CYAN}{cmyk}{1,0,0,0}
 \definecolor{MAGENTA}{cmyk}{0,1,0,0}
 \definecolor{YELLOW}{cmyk}{0,0,1,0}
}

\usepackage{graphics}\usepackage{epstopdf}\usepackage{epsfig}\@ifundefined{definecolor}{\usepackage{color}}{}
\usepackage{amsfonts}\usepackage{bm}\usepackage{amscd}

\usepackage{babel}

\makeatother

\usepackage{babel}
\begin{document}
\title{Electronic decoherence along a single nuclear trajectory }
\author{Matisse Wei-Yuan Tu}
\affiliation{Fritz Haber Center for Molecular Dynamics, Institute of Chemistry,
The Hebrew University of Jerusalem, Jerusalem 91904 Israel }
\affiliation{Max Planck Institute for the Structure and Dynamics of Matter, Luruper
Chaussee 149, 22761 Hamburg, Germany }
\author{E.K.U. Gross}
\affiliation{Fritz Haber Center for Molecular Dynamics, Institute of Chemistry,
The Hebrew University of Jerusalem, Jerusalem 91904 Israel}
\begin{abstract}
We describe a novel approach to subsystem decoherence without the
usual tracing-out of the environment. The subsystem of focus is described
entirely by a pure state evolving non-unitarily along a single classical
trajectory of its environment. The approach is deduced from the exact
factorization framework for arbitrary systems of electrons and nuclei.
The non-unitarity of the electronic dynamics arises exclusively from
non-adiabatic correlations between electrons and nuclei. We demonstrate
that the approach correctly describes the coherence gain and the subsequent
decoherence for the example of a nuclear trajectory passing through
an avoided crossing, the prototypical case where single-trajectory
Ehrenfest dynamics fails to produce decoherence. 
\end{abstract}
\maketitle

\section{Introduction}

Decoherence is a multifaceted phenomenon that appears in many branches
of physics, chemistry and biology whenever a system consists of interacting
distinct subsystems. It represents a major obstacle in the technical
realization of a scalable quantum computer. The theoretical description
starts from designating one subsystem as the primary focus, called
the principal system, while treating the rest as its ``environment''.
The wave function of the whole composite system satisfies one single
time-dependent Schr\"{o}dinger equation (TDSE). Decoherence results
from the entanglement between the principal system and its environment.
In the standard treatment \cite{Breuer02book,Chuang10book}, decoherence
effects are obtained by tracing out the environment which causes the
principal system to be in a mixed state, i.e., an ensemble average
of pure states, described by a reduced density operator (RDO). In
this work, starting from the full TDSE of the complete system, we
deduce an alternative way of tackling subsystem decoherence by propagating
just one pure state without ensemble averages. This pure state evolves
non-unitarily in a norm-conserving fashion along one classical trajectory
of its environment. The approach is derived here for arbitrary systems
of interacting electrons and nuclei, where the electronic degrees
of freedom (DOF) constitute the principal system interacting with
its nuclear environment. While this is a specific choice of two-component
system, we emphasize that the approach presented here is general and
can be applied to other composite systems in the same way \cite{Gonze18224}.

A popular family of approaches dealing with the coupled system of
electrons and nuclei are the so-called mixed quantum-classical (MQC)
approaches where the nuclear motion is described in terms of classical
trajectories while the electronic DOF are treated fully quantum mechanically.
It is usually assumed that electronic decoherence cannot be described
purely within a single trajectory, without appealing to the finite
spatial extension of nuclear wave packets. This holds true for Ehrenfest
dynamics, where the electrons evolve unitarily along a single classical
nuclear trajectory and decoherence arises only after averaging over
multiple trajectories \cite{Alonso2008096403,Xavier09728}. More broadly,
many MQC methods aim to incorporate decoherence readily on the single
trajectory level by introducing modifications, often termed decoherence
corrections, to the electronic evolution at each time step \cite{Schwartz965942,Prezhdo97825,Prezhdo975863,Hack019305,Bedard-Hearn05234106,Jasper05064103,Kaeb063197,Granucci10134111,Subotnik11024105,Subotnik16387,Esch20234105,Shu23380}.
These corrections produce deviations from the purely unitary Schrödinger
dynamics by adjusting the electronic state coefficients according
to a decoherence time parameter, which is typically determinined from
heuristically. A common rationale is the spatial separation of nuclear
wave packets on different electronic surfaces: as these wave packets
lose overlap, coherence between the corresponding electronic states
should diminish. Even energy-based decoherence-time models, which
do not explicitly track the wave packet separation, are based on in
this intuition \cite{Shu23380,Hack019305}. 

This prevailing strategy naturally leads to a curiosity: can electronic
decoherence emerge natively from a single classical trajectory, without
appealing to the finite spatial extension of nuclear wave packets?
The exact factorization (EF) formalism offers a compelling framework
in this regard \cite{Abedi10123002,Suzuki14040501,Min15073001,Li2022113001,Arribas2024233201,Abedi1222A530,Min2014263004,Requist16042108,Requist17062503}.
Briefly, the EF framework, without approximation, represents the fully
correlated wave function, $\Psi\left(\boldsymbol{r},\boldsymbol{R},t\right)$,
of interacting electrons and nuclei as a single product

\begin{equation}
\Psi\left(\boldsymbol{r},\boldsymbol{R},t\right)=\chi\left(\boldsymbol{R},t\right)\phi\left(\boldsymbol{r}\left\vert t,\boldsymbol{R}\right.\right),\label{exfc-gen1}
\end{equation}
where $\chi\left(\boldsymbol{R},t\right)$ is called the nuclear wave
function and $\phi\left(\boldsymbol{r}\left\vert t,\boldsymbol{R}\right.\right)$
an electronic wave function which parametrically depends on $\left(t,\boldsymbol{R}\right)$.
Here $\boldsymbol{r}=\left\{ \boldsymbol{r}_{j}\right\} $ and $\boldsymbol{R}=\left\{ \boldsymbol{R}_{\nu}\right\} $
denote the electronic and nuclear positions, respectively. The physical
meaning of the two factors becomes immediately clear by remembering
that the absolute square $\left\vert \Psi\left(\boldsymbol{r},\boldsymbol{R},t\right)\right\vert ^{2}$,
is the joint probability density of finding, at time $t$, the electrons
at $\boldsymbol{r}$ and the nuclei at $\boldsymbol{R}$. Being a
joint probability, $\left\vert \Psi\left(\boldsymbol{r},\boldsymbol{R},t\right)\right\vert ^{2}$
can be written as the product of a marginal and a conditional probability,
implying that $\left\vert \chi\left(\boldsymbol{R},t\right)\right\vert ^{2}$
is the marginal probability density of finding the nuclei at positions
$\boldsymbol{R}$ while $\left\vert \phi\left(\boldsymbol{r}\left\vert t,\boldsymbol{R}\right.\right)\right\vert ^{2}$
is the conditional probability density of finding the electrons at
$\boldsymbol{r}$, given the nuclei are at $\boldsymbol{R}$ at time
$t$. The nuclear factor $\chi\left(\boldsymbol{R},t\right)$, on
one hand, satisfies a standard TDSE with a time-dependent scalar potential,
$\varepsilon\left(\boldsymbol{R},t\right)$, and a time-dependent
vector potential $\boldsymbol{A}\left(\boldsymbol{R},t\right)$. The
latter has the algebraic structure of a Berry connection but, being
formally exact, it is not tied to an adiabatic approximation \cite{Min2014263004,Requist16042108,Requist17062503}.
On the other hand, the electronic factor $\phi\left(\boldsymbol{r}\left\vert t,\boldsymbol{R}\right.\right)$
satisfies a TDSE-like equation of motion (EOM) \cite{Abedi10123002,Abedi1222A530},
with a non-Hermitian Hamiltonian, i.e., its time-evolution is non-unitary
but still conserves the norm of $\phi\left(\boldsymbol{r}\left\vert t,\boldsymbol{R}\right.\right)$,
namely, $\int\text{d}\boldsymbol{r}\left\vert \phi\left(\boldsymbol{r}\left\vert t,\boldsymbol{R}\right.\right)\right\vert ^{2}$
is normalized to 1 for each fixed $\left(t,\boldsymbol{R}\right)$. 

In fact, in view of the mathematical method of characteristics \cite{Lopreore19995190,Garashchuk2010BookChp},
a set of (arbitrarily dense) classical trajectories still gives a
formally exact representation of the nuclear wave function $\chi\left(\boldsymbol{R},t\right)$.
These trajectories satisfy standard classical EOMs \cite{Talotta20206764},
and the corresponding exact forces have been identified, both in the
quantum case \cite{Li2022113001} and in the classical limit \cite{Agostini14214101}.
Since the exact-factorization of Eq. (\ref{exfc-gen1}) is a formally
exact representation of the full electron-nuclear wave function, it
should certainly describe electronic decoherence. Indeed, by treating
the nuclear dynamics with a swarm of trajectories, and combining it
with the electronic EOM of the EF (with further-going approximations
for the parametric dependence of the electronic factor on nuclear
coordinates), the resulting coupled-trajectory mixed quantum-classical
(CTMQC) algorithm was demonstrated to capture electronic decoherence,
essentially through the nuclear quantum momentum \cite{Min15073001,Min20173048,Arribas2024233201}.
On one hand, the nuclear quantum momentum explicitly derives from
the nuclear wave function, it is not intrinsic to a single classical
trajectory, which underlies the rise of coupling of trajectories.
On the other hand, a numerical difficulty associated with this algorithm
is that the coupled nature of the trajectory propagation makes it
harder to converge with the number of trajectories. While several
simplifications have led to independent-trajectory variants \cite{Ha181097,Vindel-Zandbergen213852,Ha22174109},
these developments are primarily motivated by computational efficiency
rather than by the conceptual issue we address here, namely, the nature
of decoherence when nuclear dynamics is represented by a single trajectory.

This raises two key open questions. First, although the non-unitary
yet norm-preserving evolution of the electronic conditional state
is an explicit feature of the EF formalism in its fully quantum formulation,
\cite{Abedi10123002,Suzuki14040501,Min15073001,Li2022113001,Arribas2024233201,Abedi1222A530,Min2014263004,Requist16042108,Requist17062503},
it remains unclear whether this non-unitarity persists after taking
the classical limit of the nuclei, (cf. \cite{Agostini14214101}).
To our knowledge, this has not yet been explicitly demonstrated. Second,
and more fundamentally, while non-unitarity is commonly associated
with decoherence in the context of statistical mixtures and quantum
entanglement with an environment \cite{Breuer02book,Zurek03715,Chuang10book},
its implications for pure-state, trajectory-based descriptions remain
largely unexplored. The aim of this work is to clarify how electronic
decoherence can arise within the EF formalism when the nuclear motion
is approximated by a single classical trajectory. By isolating and
analysing the internal structure of this non-unitary evolution, we
seek to identify trajectory-native factors of decoherence. In Sec.
\ref{EF-sig}, we review the core EF formalism, with particular focus
on the dynamics of the electronic factor under the classical limit
of the nuclei. In Sec. \ref{non-uni-pureStat-Ana}, we explicitly
demonstrate that the non-adiabatic correlations between electrons
and nuclei continue to dominate the non-unitarity of the electronic
subsystem\textquoteright s pure-state dynamics, even in the classical
limit for the nuclei. In Sec. \ref{non-uni-pureStat-Num}, we analyse
a prototypical scenario in which a nuclear trajectory passes through
an avoided crossing, comparing three approaches: (i) the full exact
quantum solution, (ii) mean-field Ehrenfest dynamics, and (iii) the
single-trajectory representation of EF. These comparisons reveal that
the de-orthogonalisation processes underlying the non-unitarity of
the pure electronic state dynamics can indeed account for decoherence
at the level of a single classical trajectory. Broader implications
and concluding remarks are provided in Sec. \ref{Conclu}.

\section{The single-trajectory limit of the EF}

\label{EF-sig}

We start from the Hamiltonian of a general system of interacting electrons
and nuclei, given by 
\begin{equation}
H=T_{n}+H^{BO}\left(\boldsymbol{R}\right),\label{HtnBO}
\end{equation}
where $T_{n}$ is the nuclear kinetic energy operator and $H^{BO}\left(\boldsymbol{R}\right)$
is the so-called Born-Oppenheimer (BO) Hamiltonian, 
\begin{equation}
H^{BO}\left(\boldsymbol{R}\right)=T_{e}+V_{ee}+V_{en}\left(\boldsymbol{R}\right)+V_{nn}\left(\boldsymbol{R}\right),\label{HBOe}
\end{equation}
which consists of the electronic kinetic energy $T_{e}$ and the bare
Coulomb interactions between electrons and nuclei. $H^{BO}\left(\boldsymbol{R}\right)$
acts on the electronic Hilbert space alone, and it depends parametrically
on the nuclear positions $\boldsymbol{R}$. The eigenvalues and eigenfunctions
of the BO Hamiltonian are obtained by solving
\begin{equation}
H^{BO}\left(\boldsymbol{R}\right)\left\vert \varphi_{k}\left(\boldsymbol{R}\right)\right\rangle =\varepsilon_{k}\left(\boldsymbol{R}\right)\left\vert \varphi_{k}\left(\boldsymbol{R}\right)\right\rangle ,\label{twlvl-Hilbert}
\end{equation}
at each fixed nuclear configuration $\boldsymbol{R}$. The full quantum
state $\left\vert \Psi\left(t\right)\right\rangle $ of the interacting
system of electrons and nuclei is subject to the TDSE $i\left\vert \dot{\Psi}\left(t\right)\right\rangle =H\left\vert \Psi\left(t\right)\right\rangle $.
Atomic (Hartree) units are used throughout. In terms of the solutions
of Eq. (\ref{twlvl-Hilbert}), the dynamical processes governed by
$H$ can be divided into adiabatic processes, where the expansion
of $\left\vert \Psi\left(t\right)\right\rangle $ in the complete
system of BO states is dominated by a single BO state, and non-adiabatic
processes, where transitions between BO states take place.

We characterize electronic decoherence in terms of the electronic
reduced density operator (eRDO) $\rho^{e}\left(t\right)$. The latter
is obtained by tracing out the nuclear DOF from the pure-state density
operator $\left\vert \Psi\left(t\right)\right\rangle \left\langle \Psi\left(t\right)\right\vert $
of the complete electron-nuclear system. Next we insert the exact
factorization of Eq. (\ref{exfc-gen1}) and we write the electronic
factor as $\boldsymbol{r}$-representation of an abstract $\left(t,\boldsymbol{R}\right)$-dependent
state of the electronic Hilbert space, $\phi\left(\boldsymbol{r}\left\vert t,\boldsymbol{R}\right.\right)=\left\langle \boldsymbol{r}\left\vert \phi\left(t,\boldsymbol{R}\right)\right.\right\rangle $.
This leads to 
\begin{equation}
\rho^{e}\left(t\right)=\int\text{d}\boldsymbol{R}\left\vert \chi\left(\boldsymbol{R},t\right)\right\vert ^{2}P^{e}\left(t,\boldsymbol{R}\right).\label{eRDO-f1}
\end{equation}
Here $P^{e}\left(t,\boldsymbol{R}\right)=\left\vert \phi\left(t,\boldsymbol{R}\right)\right\rangle \left\langle \phi\left(t,\boldsymbol{R}\right)\right\vert $
is the $\boldsymbol{R}$-resolved conditional electronic RDO corresponding
to the \textit{pure state} $\left\vert \phi\left(t,\boldsymbol{R}\right)\right\rangle $
while the eRDO $\rho^{e}\left(t\right)$ in general describes a mixed
state. Next we employ the classical limit \cite{Bornemann19961074,Fiete03022112,Agostini14214101}
for the nuclear wave function leading to $\left\vert \chi\left(\boldsymbol{R},t\right)\right\vert ^{2}\approx\delta\left(\boldsymbol{R}-\boldsymbol{R}_{t}^{c}\right)$
where $\boldsymbol{R}_{t}^{c}$ represents the classical nuclear configuration
at time $t$. As a consequence of this approximation, the eRDO (Eq.
(\ref{eRDO-f1})) reduces to a pure-state RDO, namely, $\rho^{e}\left(t\right)=P^{e}\left(t,\boldsymbol{R}_{t}^{c}\right)$,
evaluated along the classical trajectory $\boldsymbol{R}_{t}^{c}$.
Importantly, in contrast to Ehrenfest dynamics, a mean-field-like
factorization which removes electron-nuclear entanglement, is never
made. The electronic conditional amplitude of the exact factorization
retains electron-nuclear entanglement through its parametric dependence
on the nuclear coordinates.

Explicitly, the EOM followed by $\left\vert \phi\left(t,\boldsymbol{R}\right)\right\rangle $
reads \cite{Abedi10123002,Suzuki14040501,Min15073001,Li2022113001,Arribas2024233201,Abedi1222A530,Min2014263004,Requist16042108,Requist17062503}
\begin{subequations}\label{e-EOM-set}
\begin{equation}
i\partial_{t}\left\vert \phi\left(t,\boldsymbol{R}\right)\right\rangle =\left[H^{BO}\left(\boldsymbol{R}\right)+\mu U^{en}\left(\boldsymbol{R},t\right)\right]\left\vert \phi\left(t,\boldsymbol{R}\right)\right\rangle ,\label{e-EOM-full}
\end{equation}
where $U^{en}\left(\boldsymbol{R},t\right)=U_{K}^{en}\left[\phi\right]+U_{Q}^{en}\left[\phi,\chi\right]-\varepsilon^{na}\left[\phi\right]$
is called electron-nuclear correlation operator with
\begin{equation}
U_{K}^{en}\left[\phi\right]=\sum_{\nu}\frac{\left[-i\boldsymbol{\nabla}_{\nu}-\boldsymbol{A}_{\nu}\left[\phi\right]\left(\boldsymbol{R},t\right)\right]^{2}}{2M_{\nu}},\label{Uen_K}
\end{equation}
$\varepsilon^{na}\left[\phi\right]=\left\langle \phi\left(t,\boldsymbol{R}\right)\right\vert U_{K}^{en}\left[\phi\right]\left\vert \phi\left(t,\boldsymbol{R}\right)\right\rangle $,
and
\begin{equation}
U_{Q}^{en}\left[\phi,\chi\right]=\sum_{\nu}\boldsymbol{\mathfrak{p}}_{\nu}^{Q}\left[\phi,\chi\right]\left(\boldsymbol{R},t\right)\cdot\frac{\left[-i\boldsymbol{\nabla}_{\nu}-\boldsymbol{A}_{\nu}\left[\phi\right]\left(\boldsymbol{R},t\right)\right]}{M_{\nu}},\label{Uen_Q}
\end{equation}
in which 
\begin{equation}
\boldsymbol{\mathfrak{p}}_{\nu}^{Q}\left[\phi,\chi\right]\left(\boldsymbol{R},t\right)=\left[-i\boldsymbol{\nabla}_{\nu}\chi\left(\boldsymbol{R},t\right)/\chi\left(\boldsymbol{R},t\right)+\boldsymbol{A}_{\nu}\left[\phi\right]\left(\boldsymbol{R},t\right)\right],\label{pzero-q}
\end{equation}
\end{subequations} is the so-called nuclear momentum function with
the Berry-connection-like vector potential $\boldsymbol{A}_{\nu}\left[\phi\right]\left(\boldsymbol{R},t\right)=\left\langle \phi\left(t,\boldsymbol{R}\right)\left\vert -i\boldsymbol{\nabla}_{\nu}\phi\left(t,\boldsymbol{R}\right)\right.\right\rangle $.
The classical trajectory $\boldsymbol{R}_{t}^{c}$ of the nuclei enters
the dynamics of the pure-state electronic factor in as much as the
nuclear momentum function Eq. (\ref{pzero-q}) is taken to the classical
limit $\boldsymbol{\mathfrak{p}}_{\nu}^{Q}\left[\phi,\chi\right]\left(\boldsymbol{R},t\right)\rightarrow\boldsymbol{P}_{t}^{c}\equiv M_{\nu}\dot{\boldsymbol{R}}_{t}^{c}$
where the classical forces acting on the nuclei are given by
\begin{align}
 & M_{\nu}\ddot{R}_{\nu\alpha t}^{c}=\frac{\partial A_{\nu\alpha}\left(\boldsymbol{R}_{t}^{c},t\right)}{\partial t}-\frac{\partial\varepsilon\left(\boldsymbol{R}_{t}^{c},t\right)}{\partial R_{\nu\alpha t}^{c}}\nonumber \\
 & +\sum_{\nu^{\prime}\beta}\dot{R}_{\nu^{\prime}\beta t}^{c}\left(\frac{\partial A_{\nu\alpha}\left(\boldsymbol{R}_{t}^{c},t\right)}{\partial R_{\nu^{\prime}\beta t}^{c}}-\frac{\partial A_{\nu^{\prime}\beta}\left(\boldsymbol{R}_{t}^{c},t\right)}{\partial R_{\nu\alpha t}^{c}}\right),\label{nuc-force}
\end{align}
the electric-like (the first two terms in Eq. (\ref{nuc-force}))
and magnetic-like (the last term in the summation of Eq. (\ref{nuc-force}))
contributions \cite{Agostini14214101,Li2022113001}. Here, $\boldsymbol{R}_{t}^{c}=\left\{ R_{\nu\alpha t}^{c}\right\} $
with $\nu$ and $\alpha$ enumerating the nucleus in the system and
their spatial components and $\varepsilon\left(\boldsymbol{R},t\right)=\left\langle \phi\left(t,\boldsymbol{R}\right)\right\vert \left[H^{BO}\left(\boldsymbol{R}\right)-i\partial_{t}\right]\left\vert \phi\left(t,\boldsymbol{R}\right)\right\rangle $
is the scalar potential.

A generic difficulty in the described approach is that the (non-unitary)
propagation of the electronic conditional wave function requires knowledge
of the first- and second-order derivatives of $\phi\left(\boldsymbol{r}\left\vert t,\boldsymbol{R}\right.\right)$
with respect to the nuclear coordinates. In fact, it is exactly these
derivatives that lead to the coupling of trajectories in the CTMQC
algorithms. If $\phi\left(\boldsymbol{r}\left\vert t,\boldsymbol{R}\right.\right)$
was known exclusively along a single classical trajectory, these derivatives
could not be evaluated. To tackle this problem, we make use of the
fact that the correlated electron-nuclear problem has a small parameter,
the electronic-over-nuclear mass ratio

\begin{equation}
\mu=\frac{m_{e}}{M},\label{mass-ratio-1}
\end{equation}
where $m_{e}$ is the electron mass and $M$ is a reference nuclear
mass that can be taken to be the proton mass $m_{p}$ or the average
nuclear mass of the particular system at hand. $M_{\nu}$ is the mass
of the nucleus $\nu$ measured in units of the reference mass $M$.
Since the proton is the lightest nucleus, $\mu$ is upper bounded
by $\mu\le m_{e}/m_{p}\approx5.4\times10^{-4}$. A major advantage
of the EF is that the electronic and nuclear DOF are governed by separate
EOMs, allowing us to treat the two EOMs by different approximations.
As detailed above, the nuclear EOM is treated in classical approximation
by a single trajectory. The smallness of $\mu$ is not used there.
In fact, the limit $\mu\rightarrow0$ of the nuclear EOM leads to
static nuclei \cite{Eich16054110}, which is not a desirable starting
point for our purpose. By contrast, in Eq. (\ref{e-EOM-set}) for
the electrons, the smallness of the dimensionless $\mu$ strongly
suggests to treat $\mu U^{en}$ as a perturbation while $H^{BO}$
defines the unperturbed dynamics \cite{FootNoteVarPertMu}. The nuclear
dynamics is affected by this perturbative treatment only in as much
as the electronic wave function $\left\vert \phi\left(t,\boldsymbol{R}\right)\right\rangle $
enters the calculation of the classical forces on the nuclei. If the
electronic wave function is evaluated to a given perturbative order,
the forces on the nuclei are evaluated consistently to the same order.
We write 
\begin{equation}
\left\vert \phi\left(t,\boldsymbol{R}\right)\right\rangle =\left\vert \phi^{\left(0\right)}\left(t,\boldsymbol{R}\right)\right\rangle +\mu\left\vert \phi^{\left(1\right)}\left(t,\boldsymbol{R}\right)\right\rangle +\mathcal{O}\left(\mu^{2}\right).\label{e-expand}
\end{equation}
Based on the knowledge of the zeroth-order state $\left\vert \phi^{\left(0\right)}\left(t,\boldsymbol{R}\right)\right\rangle =e^{-itH^{BO}\left(\boldsymbol{R}\right)}\left\vert \phi\left(0,\boldsymbol{R}\right)\right\rangle $,
it is straightforward to compute the dominant non-adiabatic correction,
\begin{subequations}\label{1st-stat}
\begin{equation}
\left\vert \phi^{\left(1\right)}\left(t,\boldsymbol{R}\right)\right\rangle =-i\int_{0}^{t}\text{d}t^{\prime}e^{-i\left(t-t^{\prime}\right)H^{BO}\left(\boldsymbol{R}\right)}\left\vert \xi^{\left(0\right)}\left(t^{\prime},\boldsymbol{R}\right)\right\rangle ,\label{1st-stat1}
\end{equation}
\begin{align}
 & \left\vert \xi^{\left(0\right)}\left(t^{\prime},\boldsymbol{R}\right)\right\rangle =\left\{ U_{K}^{en}\left[\phi^{\left(0\right)}\right]+U_{Q}^{en}\left[\phi^{\left(0\right)},\chi\right]\right.\nonumber \\
 & \left.-\left\langle \phi^{\left(0\right)}\left(t^{\prime},\boldsymbol{R}\right)\right\vert U_{K}^{en}\left[\phi^{\left(0\right)}\right]\left\vert \phi^{\left(0\right)}\left(t^{\prime},\boldsymbol{R}\right)\right\rangle \right\} \left\vert \phi^{\left(0\right)}\left(t^{\prime},\boldsymbol{R}\right)\right\rangle .\label{1st-stat2}
\end{align}
\end{subequations} The pure-state electronic dynamics along a single
classical trajectory can thus be explicitly evaluated under the above
approximation described by Eqs. (\ref{e-expand}), (\ref{1st-stat})
with the involved nuclear momentum function taken to the classical
limit depicted by Eq. (\ref{nuc-force}). 

\section{non-unitary pure-state electronic dynamics}

\label{non-uni-pureStat}

\subsection{Non-unitarity and non-adiabaticity along a classical trajectory}

\label{non-uni-pureStat-Ana}

Our main interests regarding the electronic evolution specific to
a single nuclear trajectory $\boldsymbol{R}_{t}^{c}$ refers to time-dependent
pure-state processes, 
\begin{equation}
\left.\left\vert \phi\left(t,\boldsymbol{R}\right)\right\rangle \right\vert _{t=0,\boldsymbol{R}_{0}^{c}}\rightarrow\left.\left\vert \phi\left(t,\boldsymbol{R}\right)\right\rangle \right\vert _{t>0,\boldsymbol{R}_{t}^{c}}.\label{traj-bound-evoDef}
\end{equation}
In principle, such a time changing state $\left\vert \phi\left(t,\boldsymbol{R}_{t}^{c}\right)\right\rangle $
is assembled from the full knowledge of $\left\vert \phi\left(t,\boldsymbol{R}\right)\right\rangle $
for arbitrary $\boldsymbol{R}$ under the substitute of $\boldsymbol{R}$
by $\boldsymbol{R}_{t}^{c}$. The corresponding trajectory-bound electronic
evolution operator $K\left(t\right)$ is therefore defined by
\begin{equation}
\left\vert \phi\left(t,\boldsymbol{R}_{t}^{c}\right)\right\rangle =K\left(t\right)\left\vert \phi\left(0,\boldsymbol{R}_{0}^{c}\right)\right\rangle .\label{cl-eQdysProgDef}
\end{equation}
We now analyze properties of such trajectory-bound evolution operator
$K\left(t\right)$, where by construction $\boldsymbol{R}_{t}^{c}$
is taken as input.

As the general solution $\left\vert \phi\left(t,\boldsymbol{R}\right)\right\rangle $
is not available, we practically expand the evolution operator in
line with Eq. (\ref{e-expand}), namely, $K\left(t\right)=K^{\left(0\right)}\left(t\right)+\mu K^{\left(1\right)}\left(t\right)+\mathcal{O}\left(\mu^{2}\right)$
where the order-$m$ operation is defined by $\left\vert \phi^{\left(m\right)}\left(t,\boldsymbol{R}_{t}^{c}\right)\right\rangle =K^{\left(m\right)}\left(t\right)\left\vert \phi\left(0,\boldsymbol{R}_{0}^{c}\right)\right\rangle $.
The knowledge of $\left\vert \phi^{\left(m\right)}\left(t,\boldsymbol{R}\right)\right\rangle $
for $m=0,1$ for arbitrary $\boldsymbol{R}$ is available once the
initial state $\left\vert \phi\left(0,\boldsymbol{R}\right)\right\rangle $
is specified. By starting from various initial BO states $\left\vert \phi\left(0,\boldsymbol{R}\right)\right\rangle =\left\vert \varphi_{k^{\prime}}\left(\boldsymbol{R}\right)\right\rangle $,
the later-time states obtained via Eq. (\ref{1st-stat}) are re-denoted
by $\left\vert \phi^{\left(m\right)}\left(t,\boldsymbol{R}\right)\right\rangle \rightarrow\left\vert \phi_{k^{\prime}}^{\left(m\right)}\left(t,\boldsymbol{R}\right)\right\rangle $,
for $m=0,1$ with $\chi$ in Eq. (\ref{1st-stat2}) re-denoted by
$\chi_{0k^{\prime}}$. This fixes the left-hand side of Eq. (\ref{cl-eQdysProgDef})
to $\left\vert \phi\left(t,\boldsymbol{R}_{t}^{c}\right)\right\rangle =\left\{ \left\vert \phi_{k^{\prime}}^{\left(0\right)}\left(t,\boldsymbol{R}\right)\right\rangle +\mu\left\vert \phi_{k^{\prime}}^{\left(1\right)}\left(t,\boldsymbol{R}\right)\right\rangle +\mathcal{O}\left(\mu^{2}\right)\right\} _{\boldsymbol{R}=\boldsymbol{R}_{t}^{c}}$
together with $\left\vert \phi\left(0,\boldsymbol{R}_{0}^{c}\right)\right\rangle =\left\vert \varphi_{k^{\prime}}\left(\boldsymbol{R}_{0}^{c}\right)\right\rangle $
fixed on the right-hand side of Eq. (\ref{cl-eQdysProgDef}). Inserting
the complete set $\sum_{k}\left\vert \varphi_{k}\left(\boldsymbol{R}_{t}^{c}\right)\right\rangle \left\langle \varphi_{k}\left(\boldsymbol{R}_{t}^{c}\right)\right\vert $
on the right-hand side of Eq. (\ref{cl-eQdysProgDef}), the matrix
elements $\left[K\left(t\right)\right]_{lk}\equiv\left\langle \varphi_{l}\left(\boldsymbol{R}_{t}^{c}\right)\right\vert K\left(t\right)\left\vert \varphi_{k}\left(\boldsymbol{R}_{t}^{c}\right)\right\rangle $
appear. Using the knowledge of $\left\langle \varphi_{l}\left(\boldsymbol{R}\right)\left\vert \phi_{k^{\prime}}^{\left(m\right)}\left(t,\boldsymbol{R}\right)\right.\right\rangle $
for arbitrary $\boldsymbol{R}$, these matrix elements can then be
extracted upon setting $\boldsymbol{R}=\boldsymbol{R}_{t}^{c}$, assuming
that we also know $\left\langle \varphi_{k}\left(\boldsymbol{R}_{t}^{c}\right)\left\vert \varphi_{k^{\prime}}\left(\boldsymbol{R}_{0}^{c}\right)\right.\right\rangle $.
These algebra then lead to

\begin{align}
 & \left[K\left(t\right)K^{\dagger}\left(t\right)-\mathbb{I}\right]_{lk}\nonumber \\
 & =-i\mu\int_{0}^{t}\text{d}t^{\prime}e^{-i\left(t-t^{\prime}\right)\left[\varepsilon_{l}\left(\boldsymbol{R}_{t}^{c}\right)-\varepsilon_{k}\left(\boldsymbol{R}_{t}^{c}\right)\right]}\nonumber \\
 & \sum_{\nu}\frac{1}{M_{\nu}}\left\{ \left[\boldsymbol{\mathfrak{p}}_{\nu k}^{\left(0\right)Q}\left(\boldsymbol{R}_{t}^{c},t^{\prime}\right)-\boldsymbol{\mathfrak{p}}_{\nu l}^{\left(0\right)Q}\left(\boldsymbol{R}_{t}^{c},t^{\prime}\right)\right]\cdot\boldsymbol{\mathfrak{D}}_{lk}^{\nu}\left(\boldsymbol{R}_{t}^{c}\right)\right.\nonumber \\
 & +\left.\mathfrak{L}_{lk}^{\nu}\left(\boldsymbol{R}_{t}^{c}\right)-\left[\mathfrak{L}_{kl}^{\nu}\left(\boldsymbol{R}_{t}^{c}\right)\right]^{*}\right\} .\label{unitariy-checker-1r}
\end{align}
Here $\boldsymbol{\mathfrak{D}}_{lk}^{\nu}\left(\boldsymbol{R}\right)=\left\langle \varphi_{l}\left(\boldsymbol{R}\right)\right\vert \left(-i\boldsymbol{\nabla}_{\nu}-\boldsymbol{A}_{\nu}\left[\varphi_{k}\right]\left(\boldsymbol{R}\right)\right)\left\vert \varphi_{k}\left(\boldsymbol{R}\right)\right\rangle $
and $\mathfrak{L}_{lk}^{\nu}\left(\boldsymbol{R}\right)=\left\langle \varphi_{l}\left(\boldsymbol{R}\right)\right\vert \frac{1}{2}\left(-i\boldsymbol{\nabla}_{\nu}-\boldsymbol{A}_{\nu}\left[\varphi_{k}\right]\left(\boldsymbol{R}\right)\right)^{2}\left\vert \varphi_{k}\left(\boldsymbol{R}\right)\right\rangle $
are the derivative couplings of first and second order. 

Eq. (\ref{unitariy-checker-1r}) is the main result of our analysis.
The right-hand side only vanishes for $l=k$ implying that the trajectory-bound
first-order non-adiabatic evolution process is by itself non-unitary.
Note further that $\boldsymbol{\mathfrak{p}}_{\nu k}^{\left(0\right)Q}\left(\boldsymbol{R},t\right)\equiv\boldsymbol{\mathfrak{p}}_{\nu}^{Q}\left[\phi_{k}^{\left(0\right)},\chi_{0k}\right]$
is in general complex. Its imaginary part given by $-\boldsymbol{\nabla}_{\nu}\left\vert \chi_{0k}\left(\boldsymbol{R},t\right)\right\vert /\left\vert \chi_{0k}\left(\boldsymbol{R},t\right)\right\vert $
is known as the nuclear quantum momentum, which has been taken to
be the most essential factor for decoherence \cite{Min15073001,Arribas2024233201}.
Without that imaginary part, $\boldsymbol{\mathfrak{p}}_{\nu k}^{\left(0\right)Q}\left(\boldsymbol{R}_{t}^{c},t\right)$
taken along the classical trajectory $\boldsymbol{R}_{t}^{c}$ is
just the classical momentum. Importantly, even with the classical
momentum alone, the non-unitarity contributed by $\left[\boldsymbol{\mathfrak{p}}_{\nu k}^{\left(0\right)Q}\left(\boldsymbol{R}_{t}^{c},t^{\prime}\right)-\boldsymbol{\mathfrak{p}}_{\nu l}^{\left(0\right)Q}\left(\boldsymbol{R}_{t}^{c},t^{\prime}\right)\right]\cdot\boldsymbol{\mathfrak{D}}_{lk}^{\nu}\left(\boldsymbol{R}_{t}^{c}\right)$
in Eq. (\ref{unitariy-checker-1r}) is still effective. This analysis
thus clarifies that the non-unitarity persists even after taking the
classical limit of the nuclei without any consideration of nuclear
quantum momentum. We will show in the following example with explicit
numerical solution to non-unitary pure-state electronic dynamics (Eqs.
(\ref{e-expand}) and (\ref{1st-stat})) coupled to single-trajectory
classical nuclear dynamics Eq. (\ref{nuc-force}) that this non-unitarity
without nuclear quantum momentum readily produces decoherence effects.

\subsection{Pure-state non-unitarity and decoherence}

\label{non-uni-pureStat-Num}

\subsubsection{Defining coherence}

A general unitary propagation has two separate properties. On one
hand, it preserves the norm of the propagated state. On the other
hand, it keeps initially orthogonal states orthogonal. The non-unitary
approach presented here conserves the norm \cite{FootNoteNormConserv}.
Nevertheless, it leads to the de-orthogonalisation of initially orthogonal
states. In order to explore physical impacts of the abstract mathematical
non-unitarity on electronic coherence, we explicitly study two-state
systems. Then the electronic Hilbert space of interest is spanned
by two BO states $\left\{ \left\vert \varphi_{0}\left(\boldsymbol{R}\right)\right\rangle ,\left\vert \varphi_{1}\left(\boldsymbol{R}\right)\right\rangle \right\} $.
In this case, the eRDO takes the form
\begin{align}
 & \rho^{e}\left(t\right)=\int\text{d}\boldsymbol{R}\left\{ \sum_{k=0}^{1}\left\vert \chi_{k}\left(\boldsymbol{R},t\right)\right\vert ^{2}\left\vert \varphi_{k}\left(\boldsymbol{R}\right)\right\rangle \left\langle \varphi_{k}\left(\boldsymbol{R}\right)\right\vert \right.\nonumber \\
+ & \left.\left[\chi_{1}^{*}\left(\boldsymbol{R},t\right)\chi_{0}\left(\boldsymbol{R},t\right)\left\vert \varphi_{0}\left(\boldsymbol{R}\right)\right\rangle \left\langle \varphi_{1}\left(\boldsymbol{R}\right)\right\vert +\text{h.c.}\right]\right\} ,\label{eRDO-twlvl}
\end{align}
where $\left\{ \chi_{k}\left(\boldsymbol{R},t\right)\right\} $ are
the coefficients of the Born-Huang expansion, i.e., $\Psi\left(\boldsymbol{r},\boldsymbol{R},t\right)=\sum_{k}\left\langle \boldsymbol{r}\left\vert \varphi_{k}\left(\boldsymbol{R}\right)\right.\right\rangle \chi_{k}\left(\boldsymbol{R},t\right)$
\cite{Born54book}. Importantly, expanding the electronic factor of
the EF in a Born-Huang expansion as well, $\left\vert \phi\left(t,\boldsymbol{R}\right)\right\rangle =\sum_{k}C_{k}\left(t,\boldsymbol{R}\right)\left\vert \varphi_{k}\left(\boldsymbol{R}\right)\right\rangle $,
the relevant quantity in the second line of Eq. (\ref{eRDO-twlvl})
can be written as 
\begin{equation}
\chi_{1}^{*}\left(\boldsymbol{R},t\right)\chi_{0}\left(\boldsymbol{R},t\right)=\left\vert \chi\left(\boldsymbol{R},t\right)\right\vert ^{2}C_{1}^{*}\left(t,\boldsymbol{R}\right)C_{0}\left(t,\boldsymbol{R}\right).\label{BHEcprod}
\end{equation}
This will simplify the quantification of electronic coherence. In
the context of molecular physics, coherence is usually defined by
$\Gamma\left(t\right)\equiv\int\text{d}\boldsymbol{R}\chi_{1}^{*}\left(\boldsymbol{R},t\right)\chi_{0}\left(\boldsymbol{R},t\right)$
\cite{Min15073001,Arribas2024233201,Villaseco_Arribas2024054102}.
It is complex-valued in general and we call $\left\vert \Gamma\left(t\right)\right\vert $
the coherence magnitude. Approximating the nuclear density by a single
classical trajectory $\left\vert \chi\left(\boldsymbol{R},t\right)\right\vert ^{2}\approx\delta\left(\boldsymbol{R}-\boldsymbol{R}_{t}^{c}\right)$
and employing Eq. (\ref{BHEcprod}), then yields $\Gamma\left(t\right)\approx C_{1}^{*}\left(t,\boldsymbol{R}_{t}^{c}\right)C_{0}\left(t,\boldsymbol{R}_{t}^{c}\right)$.
For disambiguity, we denote the coherence computed from the above
EF-based single-trajectory approach by $\Gamma^{EF}\left(t\right)$.
To gain more insight, it is useful to compare it to another well-established
single-trajectory approach, namely, the Ehrenfest dynamics. The Ehrenfest
electronic state $\left\vert \phi^{Eh}\left(t\right)\right\rangle $
is described by $H^{BO}\left(\boldsymbol{R}_{t}^{c}\right)\left\vert \phi^{Eh}\left(t\right)\right\rangle =i\left\vert \dot{\phi}^{Eh}\left(t\right)\right\rangle $.
The coherence obtained under the Ehrenfest dynamics is then given
by $\Gamma^{Eh}\left(t\right)=\left[C_{1}^{Eh}\left(t\right)\right]^{*}C_{0}^{Eh}\left(t\right)$
with $C_{k}^{Eh}\left(t\right)=\left\langle \varphi_{k}\left(\boldsymbol{R}_{t}^{c}\right)\left\vert \phi^{Eh}\left(t\right)\right.\right\rangle $. 

\subsubsection{Messages from the exact solutions of the full problem}

To be concrete, we focus on a widely studied representative scenario,
in which the electronic coherence $\left\vert \Gamma\left(t\right)\right\vert $
rises and falls (decoherence) along with the passage of the nuclear
wave packet through an avoided crossing \cite{Abedi2013263001,Min15073001}.
We take a simple modelling of the avoided crossing by $H^{BO}\left(R\right)=R\left(\Delta/2\right)\sigma_{z}+g\sigma_{x}$.
Here $R$ is the nuclear coordinate and $\sigma_{x},\sigma_{z}$ are
the Pauli matrices of the two-state electronic Hilbert space. The
parameters $g$ and $\Delta$ determine the shape of the avoided crossing.
The exact numerical solution to the TDSE of the full problem for this
model is available. For the typical scenario of studying the dynamics
of electronic coherence along passing the nuclear wave packet through
the avoided crossing, the initial state is such that the nuclear density
is a Gaussian with the electrons sitting on the upper BO surface \cite{FootNoteUnitSet}.

\begin{figure}[h] \includegraphics[width=9cm, height=5.0 cm]{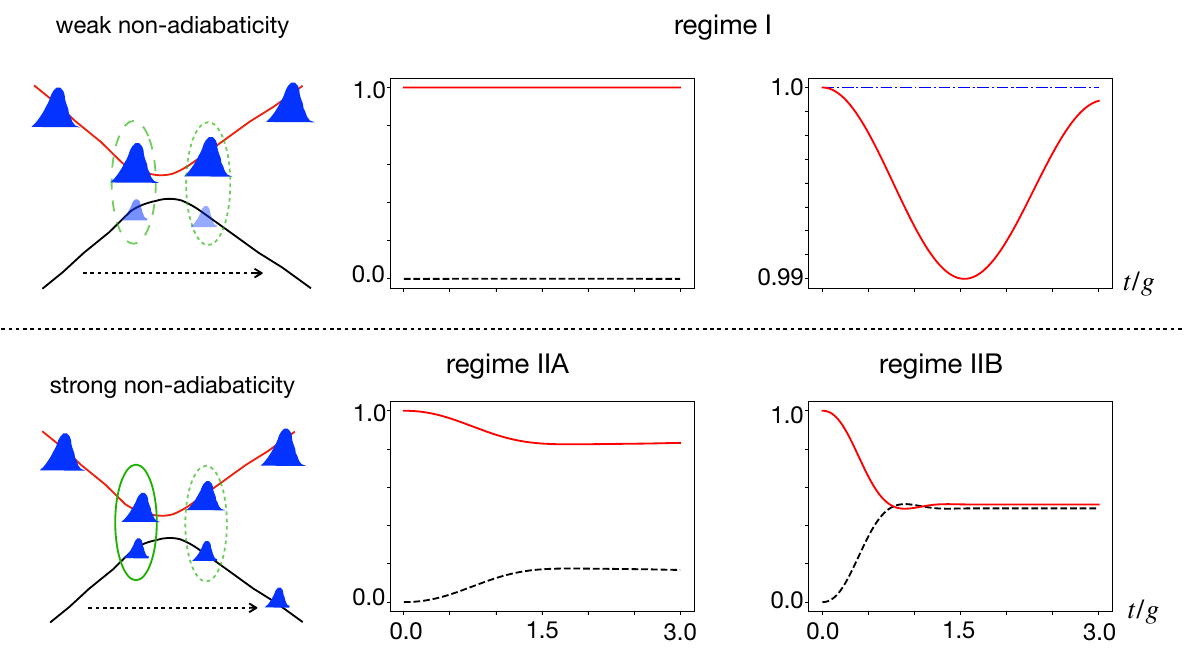} 
\caption{Distinct regimes exhibited by the avoided-crossing system used here.  Upper panel: the regime of weak non-adiabatic transition (regime I).  The inter-BO-surface population transfer is almost negligible (red-solid/black dashed for upper/lower-surface population).  The blue dashdot line in the upper right plot indicates unity population for eye guide.  Lower panel: strong non-adiabatic transitions with moderate transfer (regime IIA) and nearly 50$\%$ transfer (regime IIB).  Here we demonstrate these regimes by setting the model parameters to  $M=20$, $\Delta=1.0$, $P^{c}_{0}=1.04$, $R^{c}_{0}=-0.1$ for Regime I and  $M=15$,$\Delta=3.0$, $P^{c}_{0}=15$, $R^{c}_{0}=-1$ Regime IIA while Regime IIB uses $M=5$ with other parameters being the same as Regime IIA. } 
\label{regimes} 
\end{figure}

To deepen our understanding of the mechanisms behind electronic decoherence,
we now turn to exact numerical solutions of the full electron-nuclear
dynamics across three qualitatively distinct non-adiabatic regimes,
overviewed in Fig. \ref{regimes}. These regimes differ primarily
in the extent of population transfer between the Born-Oppenheimer
(BO) electronic states: Regime I corresponds to negligible transfer
($<1\%$), Regime IIA to modest but non-negligible transfer (sub-50\%),
and Regime IIB to strong transfer (approaching or exceeding 50\%).
The correlation between non-adiabaticity and non-unitarity, addressed
analytically through Eq. (\ref{unitariy-checker-1r}) above, will
be made more concrete in terms of decoherence later.

\begin{figure}[h] \includegraphics[width=9cm, height=7.5 cm]{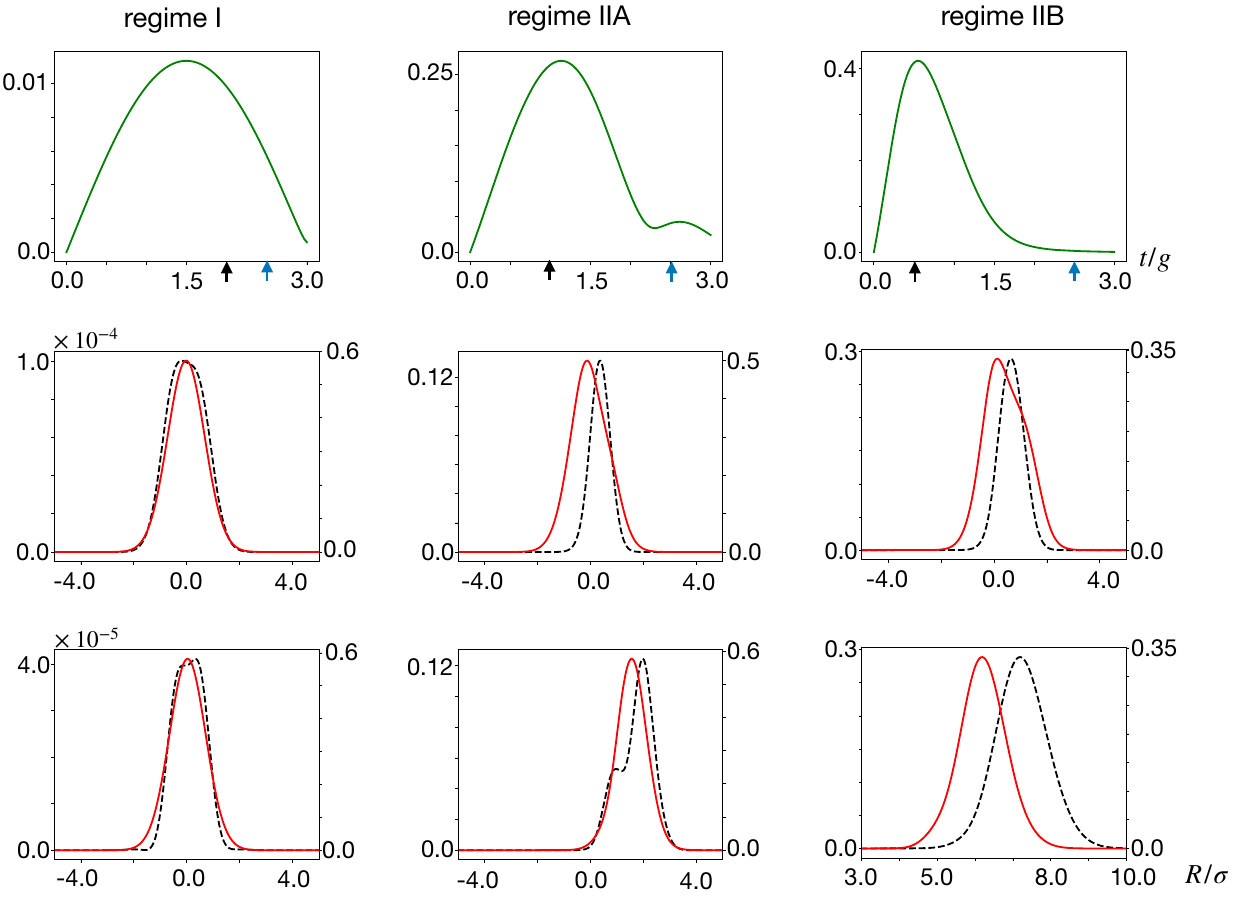} 
\caption{Coherence magnitude $\left\vert\Gamma(t) \right\vert$ are shown by solid green lines on the first row. The second/third row takes snapshots of nuclear densities at the time indicated by the black/blue arrow on the time axis of the first row.  The upper-surface densities $\left\vert\chi_{1}(t) \right\vert^{2}$ are red solid curves calibrated by the right vertical axis and the lower-surface densities $\left\vert\chi_{0}(t) \right\vert^{2}$ are black dashed lines calibrated by the left-vertical axis. } 
\label{coh_surfDens} 
\end{figure}

In Fig. \ref{coh_surfDens}, we examine the dynamics of coherence
and associated nuclear densities on the upper ($\left\vert \chi_{1}\left(R,t\right)\right\vert ^{2}$)
and the lower ($\left\vert \chi_{0}\left(R,t\right)\right\vert ^{2}$)
BO surfaces in these regimes. The first row of this figure plots the
exact coherence magnitude $\left\vert \Gamma\left(t\right)\right\vert $
over time for each regime. While all three exhibit a pattern of rise
and fall, the physical mechanisms underlying these temporal profiles
actually differ and are more deeply revealed in the snapshots of $\left\vert \chi_{1/0}\left(R,t\right)\right\vert ^{2}$
shown in the second and third rows, taken respectively at the instances
with contrasted large and small magnitudes of coherence. 

In Regime I (left column), the observed rise and fall of coherence
is strongly correlated with population dynamics: a small fraction
of the upper population is temporarily transferred to the lower BO
surface and then returns to the upper surface. So the product $\left\vert \chi_{1}\left(R,t\right)\chi_{0}\left(R,t\right)\right\vert $
naturally follows this pattern of rise and fall. The second and third
row plots show that the region in which the density $\left\vert \chi_{1}\left(R,t\right)\right\vert ^{2}$
is nonzero overlaps very much with that for $\left\vert \chi_{0}\left(R,t\right)\right\vert ^{2}$
at the moment where the coherence is high (middle row) as well as
to the moment the coherence has decayed to nearly zero (the last row).
In this weakly coupled regime, the apparent decay of coherence is
a direct consequence of reversible population transfer. The relative
phases between $\chi_{1}\left(R,t\right)$ and $\chi_{0}\left(R,t\right)$
as well as the time-changing overlap between the two wave packets,
as the most commonly attributed factors behind decoherence, are not
essential in this regime.

In contrast, Regime IIB (right column) showcases a more canonical
case of electronic decoherence. Here, as the coherence temporarily
becomes smaller, the nuclear wave packets on different surfaces also
temporarily become spatially separated, (see the changes from the
second row at higher coherence to the last row with diminished coherence).
This aligns with the commonly invoked physical picture of decoherence
due to spatial separation and diminishing overlap between nuclear
wave packets.

Interestingly, Regime IIA (middle column) presents a more nuanced
scenario. The coherence decay in this case is not associated with
visible spatial separation of the wave packets. In fact, comparing
the snapshots for Regime IIA, the spatial overlap between $\chi_{0}\left(R,t\right)$
and $\chi_{1}\left(R,t\right)$ is greater at the later time (low
coherence) than at the earlier time (high coherence). Hence, the reduction
of coherence here cannot be attributed to loss of spatial overlap.
Though the integrand in $\Gamma\left(t\right)$ remains locally nonzero,
the overall integral vanishes due to cancellation across $R$. The
well-established dephasing/phase jittering effects are consistent
with this observation \cite{Fiete03022112,Vacher15040502}. 

Altogether, the insights from Fig. \ref{coh_surfDens} establish that
the observed coherence decay can result from multiple distinct mechanisms.
These include reversible population transfer (Regime I), cancellation
of contributions across $R$ (Regime IIA) as well as nuclear wave
packet separation (Regime IIB). These findings motivate the subsequent
exploration of decoherence factors intrinsic to the single-trajectory
dynamics of the electronic subsystem that are not directly identifiable
from examining the BO-surface-resolved nuclear densities. This sets
the stage for our main inquiry in the following subsection.

\subsubsection{Single-trajectory factors behind the dynamics of coherence}

\begin{figure}[h] \includegraphics[width=9cm, height=7.5 cm]{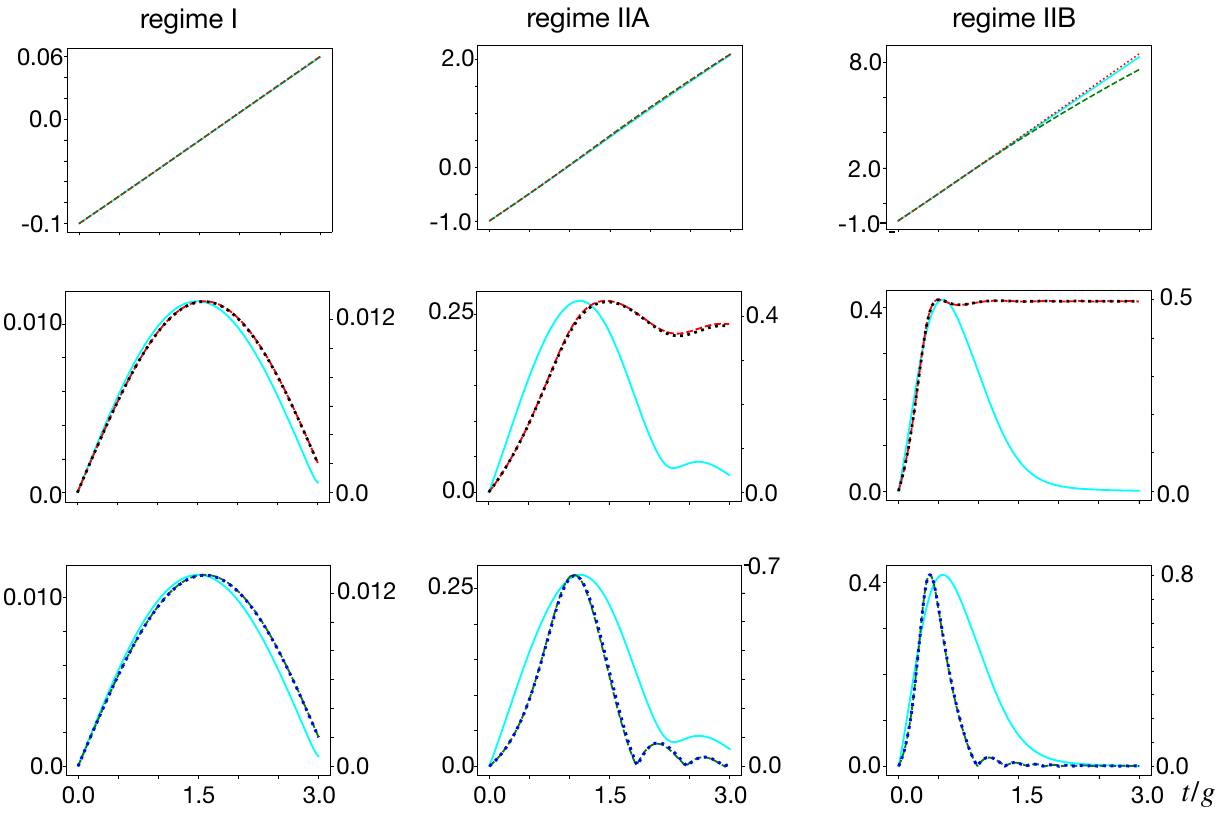} 
\caption{The first row show trajectories from exact solution (cyan solid), EF-based single trajectory method (green dashed) and the Ehrenfest method (red dotted).  The exact coherence dynamics is replicated by cyan solid lines in the second and the third row.  The Ehrenfest coherence with self-consistently obtained trajectories of the first row is shown as the red dashed lines while that obtained by input exact trajectories is shown as the black dotted lines.  The third row shows coherence from the EF-based method with self-consistently obtained trajectories (green dashed) and with exact trajectories (blue dotted). Despite slight deviations in these two trajectories in Regime IIB, the coherence from the EF-based method associated with these two trajectories agree with each other.} 
\label{coh_traj} 
\end{figure}

Having identified multiple underlying causes for coherence decay in
exact quantum dynamics, we now examine to which extent such phenomenon
can be captured using single-trajectory methods. Specifically, we
compare the coherence dynamics obtained from three approaches: (i)
the EF-based single-trajectory scheme developed in this work ($\Gamma^{EF}\left(t\right)=C_{1}^{*}\left(t,\boldsymbol{R}_{t}^{c}\right)C_{0}\left(t,\boldsymbol{R}_{t}^{c}\right)$),
(ii) the Ehrenfest method ( $\Gamma^{Eh}\left(t\right)=\left[C_{1}^{Eh}\left(t\right)\right]^{*}C_{0}^{Eh}\left(t\right)$),
and (iii) the exact quantum dynamics used as a reference benchmark
$\Gamma\left(t\right)=\int\text{d}\boldsymbol{R}\chi_{1}^{*}\left(\boldsymbol{R},t\right)\chi_{0}\left(\boldsymbol{R},t\right)$.
Fig. \ref{coh_traj} summarises these comparisons across the same
three regimes discussed previously.

The exact solution calculates the trajectory as the expectation value
of the nuclear coordinate, namely, $\boldsymbol{R}_{t}^{c}=\int\text{d}\boldsymbol{R}\left\vert \chi\left(\boldsymbol{R},t\right)\right\vert ^{2}\boldsymbol{R}=\int\text{d}\boldsymbol{R}\left\vert \chi_{0}\left(\boldsymbol{R},t\right)\right\vert ^{2}\boldsymbol{R}+\int\text{d}\boldsymbol{R}\left\vert \chi_{1}\left(\boldsymbol{R},t\right)\right\vert ^{2}\boldsymbol{R}$.
The first row of Fig. \ref{coh_traj} demonstrates near-perfect agreement
with results obtained from trajectories computed within each respective
method. This indicates that the classical approximations on the nuclei
from the single-trajectory methods (i) and (ii) are appropriate. On
the right column of Fig. \ref{coh_traj}, a slight deviation of the
trajectory given by method (i) from the exact solution is observed.
To see how such deviations from the exact classical trajectories affect
coherence dynamics, in addition to computing coherence using methods
(i) and (ii) coupled to self-consistently obtained trajectories within
those methods themselves, we also input the exact trajectories $R_{t}^{c}$
to the electronic propagations of these methods to obtain the electronic
coherence (see captions for the second and the third rows of Fig.
\ref{coh_traj}).

We now discuss the subsequent the coherence dynamics. The second and
third rows of Fig. \ref{coh_traj} plot the results from the Ehrenfest
and EF-based methods, respectively. Each plot includes the exact coherence
(cyan solid line) for direct comparison. The Ehrenfest results in
Regime I are close to the exact coherence. In Regimes IIA and IIB,
however, the Ehrenfest coherence exhibits a clear failure: after reaching
its peak, it decays only partially and then plateaus at a stable nonzero
value, in contrast to the complete or near-complete decay seen in
the exact solution. This discrepancy persists whether one uses the
self-consistent Ehrenfest trajectory (the red-dashed line) or the
exact nuclear trajectory as input (the black dotted line). This result
reinforces the well-known failure of Ehrenfest dynamics in capturing
electronic decoherence \cite{Tully98407,Makhov14054110,Fedorov194542,Esch21214101}.

The EF-based single-trajectory method performs differently. While
it overestimates the peak magnitude of coherence in Regimes IIA and
IIB, it captures the subsequent decay behavior correctly, namely,
the coherence decays to nearly zero on a timescale comparable to that
of the exact result. The deviation in classical trajectory of the
EF-based method from the exact solution in Regime IIB shows no visible
effects on the coherence dynamics, the results in coherence computed
from self-consistently obtained classical trajectories (green dashed
lines) and that from taking the exact nuclear trajectory as input
(blue dotted lines) agree with each other. These observations indicate
that something intrinsic to the structure of the EF-based single-trajectory
electronic dynamics---namely, its non-unitarity---accounts for the
observed decoherence. This conclusion is reinforced by the final row
of Fig. \ref{coh_traj}, which shows that in moving from Regime IIA
to IIB, the increasing non-adiabatic transition is accompanied by
increased rapidity of losing coherence (higher coherence magnitude
is reached earlier in time and decays faster to zero in Regime IIB
than Regime IIA). This echos with the analysis in Sec. IIIA which
shows that non-adiabatic transitions are responsible for driving non-unitarity
in the electronic evolution that is responsible for decoherence.

That the unitary Ehrenfest dynamics fails to reproduce coherence decay,
while the non-unitary EF-based method succeeds, strongly suggests
that non-unitarity triggered by non-adiabatic transitions is the trajectory-native
factor behind decoherence. We therefore have verified that the structure
of the EF electronic dynamics along a single nuclear classical trajectory
inherently encodes decoherence mechanisms, independent of ensemble
averaging or environmental entanglement. This constitutes the central
result of our investigation.

\section{Conclusions}

\label{Conclu}

In many mixed quantum--classical approaches, decoherence along a
single nuclear trajectory is introduced by manually resetting the
electronic states at each time step \cite{Schwartz965942,Prezhdo97825,Prezhdo975863,Hack019305,Bedard-Hearn05234106,Jasper05064103,Kaeb063197,Granucci10134111,Subotnik11024105,Subotnik16387,Esch20234105,Shu23380}.
These decoherence corrections make the electronic evolution effectively
non-unitary over each time step, in line with the general understanding
that decoherence is associated with departures from unitarity \cite{Zurek03715}.
However, the form and application of such corrections are typically
heuristic, relying on intuitive models that utilize the finite widths
of the wave packets to determine how coherence should be reduced \cite{Schwartz965942,Prezhdo97825,Prezhdo975863,Hack019305,Bedard-Hearn05234106,Jasper05064103,Kaeb063197,Granucci10134111,Subotnik11024105,Subotnik16387,Esch20234105,Shu23380}. 

By contrast, the non-unitary evolution investigated in this work emerges
intrinsically from the structure of the EF formalism, even when the
nuclear degrees of freedom are described by a single classical trajectory.
No a priori resetting procedure is required, nor is there any reliance
on the spatial widths of nuclear wave packets. Instead, non-adiabatic
transitions directly induce non-unitarity in the dynamics of the electronic
conditional state, which in turn leads to the decay of electronic
coherence---capturing decoherence natively at the single-trajectory
level.

We expect future applications to periodic solids where the Born-Huang
expansion is not feasible in practice, while our perturbative approach
is readily applicable, especially when the electronic structure part
is treated with the EF version of density functional theory \cite{Requist2016193001,Requist2019165136}.
Finally, the approach can be easily generalized to other composite
systems since the EF framework is valid for arbitrary two- or multi-component
systems \cite{Gonze18224}. 

\section*{Acknowledgement}

We thank Aaron Kelly, Jonathan Mannouch, Sebastian de la Pena for
useful discussions. This project has received funding from the European
Research Council (ERC) under the European Union's Horizon 2020 research
and innovation programme (grant agreement No. ERC-2017-AdG-788890).
E.K.U.G. acknowledges support as Mercator fellow within SFB 1242 at
the University Duisburg-Essen.

\bibliographystyle{myunsrt} 
\bibliography{refs_bibexd-1} 

\begin{thebibliography}{10}
\bibitem{FootNoteVarPertMu}
There are several possibilities to develop perturbation theories based on this
  small parameter, differing by the choice of unperturbed limit and the
  physical processes to be treated
  \cite{Born27457,Hagedorn86571,Nafie925687,Scherrer135305,Scherrer15074106,Eich16054110,
  Axel163316,Scherrer17031035}.
\bibitem{FootNoteNormConserv}
This can be seen straightforwardly by examining
  $\partial_{t}\text{\ensuremath{\left(\left\langle
  \phi\left(t,\boldsymbol{R}\right)\left\vert
  \phi\left(t,\boldsymbol{R}\right)\right.\right\rangle \right)}}$ using Eq.
  (\ref{e-EOM-set}) for the full electronic factor $\left\vert
  \phi\left(t,\boldsymbol{R}\right)\right\rangle$. Alternatively, one can
  substitute the expansion of Eq. (\ref{e-expand}) to arbitrary order $m\ge1$
  to $\partial_{t}\text{\ensuremath{\left(\left\langle
  \phi\left(t,\boldsymbol{R}\right)\left\vert
  \phi\left(t,\boldsymbol{R}\right)\right.\right\rangle \right)}}$ and yield
  $\sum_{n=0}^{m}\left\langle
  \phi^{\left(n\right)}\left(t,\boldsymbol{R}\right)\left\vert
  \phi^{\left(m-n\right)}\left(t,\boldsymbol{R}\right)\right.\right\rangle =0$.
  This result upon taking $m=1$ translates to $\left\langle
  \phi\left(t,\boldsymbol{R}\right)\left\vert
  \phi\left(t,\boldsymbol{R}\right)\right.\right\rangle
  =1+\mathcal{O}\left(\mu^{2}\right)$, which can also be verified explicitly
  through Eq. (\ref{1st-stat}). Notably, these results hold regardless of how
  one treats the nuclear factor $\chi$.
\bibitem{FootNoteUnitSet}
Since the actual electron mass $m_{e}$ as well as the associated Bohr radius
  evoked in the atomic units used for general analysis is not really involved
  in models like the one exemplified here, for numerical calculations, it is
  more convenient to take the units of energy/time and coordinate/momentum
  directly from the defining parameters of the model. It is natural to take $g$
  as the energy unit. Noticeably, when $\Delta=0$, then the non-adiabatic
  coupling $d\left(R\right)\equiv\left\langle
  \varphi_{0}\left(R\right)\left\vert
  \partial\varphi_{1}\left(R\right)/\partial R\right.\right\rangle$ vanishes
  identically for all $R$'s. In order to compare situations of various
  strengths of non-adiabatic couplings, one should allow $\Delta$ to vary.
  Therefore, though $g/\Delta$ has the physical dimension of length, it is not
  used for the unit of length. The situation $\Delta=0$ corresponds to that the
  nuclear particle is a free particle with the internal electronic states fixed
  to one of the two eigenstates of $\sigma_{x}$, independent of $R$. As a free
  particle, the initial Gaussian width $\sigma$ conveniently serves as the
  length unit which can still serve the purpose well for $\Delta\ne0$. Then
  $\Delta$ is measured in unit of $g/\sigma$ and the nuclear mass M is in unit
  of $\sigma^{-2}/g$ (with $\hbar$=1).
\end{thebibliography}
\end{document}